\documentstyle[twoside]{article}

\catcode`\@=11
\long\def\@makefntext#1{
\protect\noindent \hbox to 3.2pt {\hskip-.9pt  
$^{{\eightrm\@thefnmark}}$\hfil}#1\hfill}		

\def\@makefnmark{\hbox to 0pt{$^{\@thefnmark}$\hss}}	
	
\def\ps@myheadings{\let\@mkboth\@gobbletwo
\def\@oddhead{\hbox{}
\rightmark\hfil\eightrm\thepage}   
\def\@oddfoot{}\def\@evenhead{\eightrm\thepage\hfil
\leftmark\hbox{}}\def\@evenfoot{}
\def\sectionmark##1{}\def\subsectionmark##1{}}

\oddsidemargin=\evensidemargin
\newcounter{sectionc}\newcounter{subsectionc}\newcounter{subsubsectionc}
\renewcommand{\section}[1] {\vspace{12pt}\addtocounter{sectionc}{1} 
\setcounter{subsectionc}{0}\setcounter{subsubsectionc}{0}\noindent 
	{\tenbf\thesectionc. #1}\par\vspace{5pt}}
\renewcommand{\subsection}[1] {\vspace{12pt}\addtocounter{subsectionc}{1} 
	\setcounter{subsubsectionc}{0}\noindent 
	{\bf\thesectionc.\thesubsectionc. {\kern1pt \bfit #1}}\par\vspace{5pt}}
\renewcommand{\subsubsection}[1] {\vspace{12pt}\addtocounter{subsubsectionc}{1}
	\noindent{\tenrm\thesectionc.\thesubsectionc.\thesubsubsectionc.
	{\kern1pt \tenit #1}}\par\vspace{5pt}}
\newcommand{\nonumsection}[1] {\vspace{12pt}\noindent{\tenbf #1}
	\par\vspace{5pt}}

\newcounter{appendixc}
\newcounter{subappendixc}[appendixc]
\newcounter{subsubappendixc}[subappendixc]
\renewcommand{\thesubappendixc}{\Alph{appendixc}.\arabic{subappendixc}}
\renewcommand{\thesubsubappendixc}
	{\Alph{appendixc}.\arabic{subappendixc}.\arabic{subsubappendixc}}

\renewcommand{\appendix}[1] {\vspace{12pt}
        \refstepcounter{appendixc}
        \setcounter{figure}{0}
        \setcounter{table}{0}
        \setcounter{lemma}{0}
        \setcounter{theorem}{0}
        \setcounter{corollary}{0}
        \setcounter{definition}{0}
        \setcounter{equation}{0}
        \renewcommand{\thefigure}{\Alph{appendixc}.\arabic{figure}}
        \renewcommand{\thetable}{\Alph{appendixc}.\arabic{table}}
        \renewcommand{\theappendixc}{\Alph{appendixc}}
        \renewcommand{\thelemma}{\Alph{appendixc}.\arabic{lemma}}
        \renewcommand{\thetheorem}{\Alph{appendixc}.\arabic{theorem}}
        \renewcommand{\thedefinition}{\Alph{appendixc}.\arabic{definition}}
        \renewcommand{\thecorollary}{\Alph{appendixc}.\arabic{corollary}}
        \renewcommand{\theequation}{\Alph{appendixc}.\arabic{equation}}
        \noindent{\tenbf Appendix \theappendixc #1}\par\vspace{5pt}}
\newcommand{\subappendix}[1] {\vspace{12pt}
        \refstepcounter{subappendixc}
        \noindent{\bf Appendix \thesubappendixc. {\kern1pt \bfit #1}}
	\par\vspace{5pt}}
\newcommand{\subsubappendix}[1] {\vspace{12pt}
        \refstepcounter{subsubappendixc}
        \noindent{\rm Appendix \thesubsubappendixc. {\kern1pt \tenit #1}}
	\par\vspace{5pt}}

\newcommand{\textlineskip}{\baselineskip=13pt}
\newcommand{\smalllineskip}{\baselineskip=10pt}

\def\eightcirc{
\begin{picture}(0,0)
\put(4.4,1.8){\circle{6.5}}
\end{picture}}
\def\eightcopyright{\eightcirc\kern2.7pt\hbox{\eightrm c}} 

\newcommand{\copyrightheading}[1]
	{\vspace*{-2.5cm}\smalllineskip{\flushleft
	{\footnotesize International Journal of Modern Physics A, #1}\\
	{\footnotesize $\eightcopyright$\, World Scientific Publishing
	 Company}\\
	 }}

\def\abstracts#1#2#3{{
	\centering{\begin{minipage}{4.5in}\baselineskip=10pt\footnotesize
	\parindent=0pt #1\par 
	\parindent=15pt #2\par
	\parindent=15pt #3
	\end{minipage}}\par}} 

\renewenvironment{thebibliography}[1]
	{\frenchspacing
	 \ninerm\baselineskip=11pt
	 \begin{list}{\arabic{enumi}.}
	{\usecounter{enumi}\setlength{\parsep}{0pt}
	 \setlength{\leftmargin 12.7pt}{\rightmargin 0pt} 
	 \setlength{\itemsep}{0pt} \settowidth
	{\labelwidth}{#1.}\sloppy}}{\end{list}}

\newcounter{itemlistc}
\newcounter{romanlistc}
\newcounter{alphlistc}
\newcounter{arabiclistc}

\newcommand{\fcaption}[1]{
        \refstepcounter{figure}
        \setbox\@tempboxa = \hbox{\footnotesize Fig.~\thefigure. #1}
        \ifdim \wd\@tempboxa > 5in
           {\begin{center}
        \parbox{5in}{\footnotesize\smalllineskip Fig.~\thefigure. #1}
            \end{center}}
        \else
             {\begin{center}
             {\footnotesize Fig.~\thefigure. #1}
              \end{center}}
        \fi}

\newcommand{\tcaption}[1]{
        \refstepcounter{table}
        \setbox\@tempboxa = \hbox{\footnotesize Table~\thetable. #1}
        \ifdim \wd\@tempboxa > 5in
           {\begin{center}
        \parbox{5in}{\footnotesize\smalllineskip Table~\thetable. #1}
            \end{center}}
        \else
             {\begin{center}
             {\footnotesize Table~\thetable. #1}
              \end{center}}
        \fi}

\def\@citex[#1]#2{\if@filesw\immediate\write\@auxout
	{\string\citation{#2}}\fi
\def\@citea{}\@cite{\@for\@citeb:=#2\do
	{\@citea\def\@citea{,}\@ifundefined
	{b@\@citeb}{{\bf ?}\@warning
	{Citation `\@citeb' on page \thepage \space undefined}}
	{\csname b@\@citeb\endcsname}}}{#1}}

\newif\if@cghi
\def\cite{\@cghitrue\@ifnextchar [{\@tempswatrue
	\@citex}{\@tempswafalse\@citex[]}}
\def\citelow{\@cghifalse\@ifnextchar [{\@tempswatrue
	\@citex}{\@tempswafalse\@citex[]}}
\def\@cite#1#2{{$\null^{#1}$\if@tempswa\typeout
	{IJCGA warning: optional citation argument 
	ignored: `#2'} \fi}}

\def\pmb#1{\setbox0=\hbox{#1}
	\kern-.025em\copy0\kern-\wd0
	\kern.05em\copy0\kern-\wd0
	\kern-.025em\raise.0433em\box0}

\def\fnt#1#2{\footnotetext{\kern-.3em
	{$^{\mbox{\scriptsize #1}}$}{#2}}}

\def\fpage#1{\begingroup
\voffset=.3in
\thispagestyle{empty}\begin{table}[b]\centerline{\footnotesize #1}
	\end{table}\endgroup}

\def\runninghead#1#2{\pagestyle{myheadings}
\markboth{{\protect\footnotesize\it{\quad #1}}\hfill}
{\hfill{\protect\footnotesize\it{#2\quad}}}}
\font\tenrm=cmr10
\font\tenit=cmti10 
\font\tenbf=cmbx10
\font\bfit=cmbxti10 at 10pt
\font\ninerm=cmr9

\font\eightrm=cmr8

\def\qed{\hbox{${\vcenter{\vbox{			
   \hrule height 0.4pt\hbox{\vrule width 0.4pt height 6pt
   \kern5pt\vrule width 0.4pt}\hrule height 0.4pt}}}$}}

\begin{document}

\input{psfig}

\def\VEV#1{{\left\langle #1 \right\rangle}}
\def\lsim{\mathrel{\rlap{\lower4pt\hbox{\hskip1pt$\sim$}}
    \raise1pt\hbox{$<$}}}         
\def\gsim{\mathrel{\rlap{\lower4pt\hbox{\hskip1pt$\sim$}}
    \raise1pt\hbox{$>$}}}         
\def\hatn{{\bf \hat n}}

\runninghead{Detection of Gravitational Waves from Inflation}{Detection of Gravitational Waves from Inflation}

\normalsize\textlineskip
\thispagestyle{empty}
\setcounter{page}{1}

\copyrightheading{}			

\vspace*{0.88truein}

\fpage{1}
\centerline{\bf DETECTION OF GRAVITATIONAL WAVES FROM INFLATION}
\vspace*{0.37truein}
\centerline{\footnotesize MARC KAMIONKOWSKI\footnote{kamion@tapir.caltech.edu}}
\vspace*{0.015truein}
\centerline{\footnotesize\it California Institute of Technology, Mail Code
130-33}
\baselineskip=10pt
\centerline{\footnotesize\it Pasadena, CA~~91125, USA}
\vspace*{10pt}
\centerline{\footnotesize ANDREW H. JAFFE\footnote{jaffe@cfpa.berkeley.edu}}
\vspace*{0.015truein}
\centerline{\footnotesize\it Center for Particle Astrophysics,
301 LeConte Hall, University of California}
\baselineskip=10pt
\centerline{\footnotesize\it Berkeley, CA~~94720, USA}

\vspace*{0.21truein}
\abstracts{Recent measurements of temperature fluctuations in the cosmic
microwave background (CMB) indicate that the Universe is flat and that
large-scale structure grew via gravitational infall from primordial
adiabatic perturbations.  Both of these observations seem to
indicate that we are on the right track with inflation.  But
what is the new physics responsible for inflation?  This
question can be answered with observations of the polarization of the CMB.
Inflation predicts robustly the existence of a stochastic
background of cosmological gravitational waves with an amplitude
proportional to the square of the energy scale of inflation.
This gravitational-wave background induces a unique signature in 
the polarization of the CMB.  If inflation took place at an
energy scale much smaller than that of grand unification, then
the signal will be too small to be detectable.  However,
if inflation had something to do with grand unification or
Planck-scale physics, then the signal is conceivably
detectable in the optimistic case by the Planck satellite, or if 
not, then by a dedicated post-Planck CMB polarization experiment.
Realistic developments in detector technology as well as a
proper scan strategy could produce such a post-Planck experiment
that would improve on Planck's sensitivity to the
gravitational-wave background by several orders of magnitude in a
decade timescale.}{}{}


\vspace*{1pt}\textlineskip	
\section{What Have We Learned from the Cosmic Microwave Background?}
\vspace*{-0.5pt}
\noindent
The past year has seen spectacular advances in measurements of
temperature fluctuations in the cosmic microwave background
(CMB)\cite{Miletal99,deBetal00,Hanetal00} that have led to
major advances in our ability to 
characterize the largest-scale structure of the Universe, the
origin of density perturbations, and the early Universe.  The
primary aim of these experiments has been to determine the power 
spectrum, $C_\ell$, of the CMB as a function of multipole moment 
$\ell$.  Given a map of the temperature $T(\hatn)$ in each 
direction $\hatn$ on the sky, the power spectrum can be
obtained by expanding in spherical harmonics,
\begin{equation}
     a_{\ell m} = \int \, d\hatn\, Y_{\ell m}(\hatn)
     \, T(\hatn),
\end{equation}
and then squaring and summing the coefficients,
\begin{equation}
     C_\ell = {1\over 2\ell +1} \sum_m |a_{\ell m}|^2.
\end{equation}
If the map covers a patch of the sky that is small enough to be
approximated as a flat surface, the power spectrum can be
written in terms of Fourier coefficients:
\begin{equation}
     T_{\vec\ell} = \int \, d\hatn\, e^{-i \vec \ell \cdot
     \vec \theta}\, T(\hatn),
\end{equation}
and then
\begin{equation}
     C_\ell \simeq \VEV{|T_{\vec\ell}|^2}_{|\ell|=\ell},
\end{equation}
where the average is taken over all Fourier coefficients $\vec
\ell$ that have amplitude $\ell$.  Thus, each multipole moment
$C_\ell$ measures, roughly speaking, the rms temperature
fluctuation between two points separated by an angle $\theta
\simeq (\ell/200)^{-1}$ degrees on the sky.

\begin{figure}[tbp]
\centerline{\psfig{file=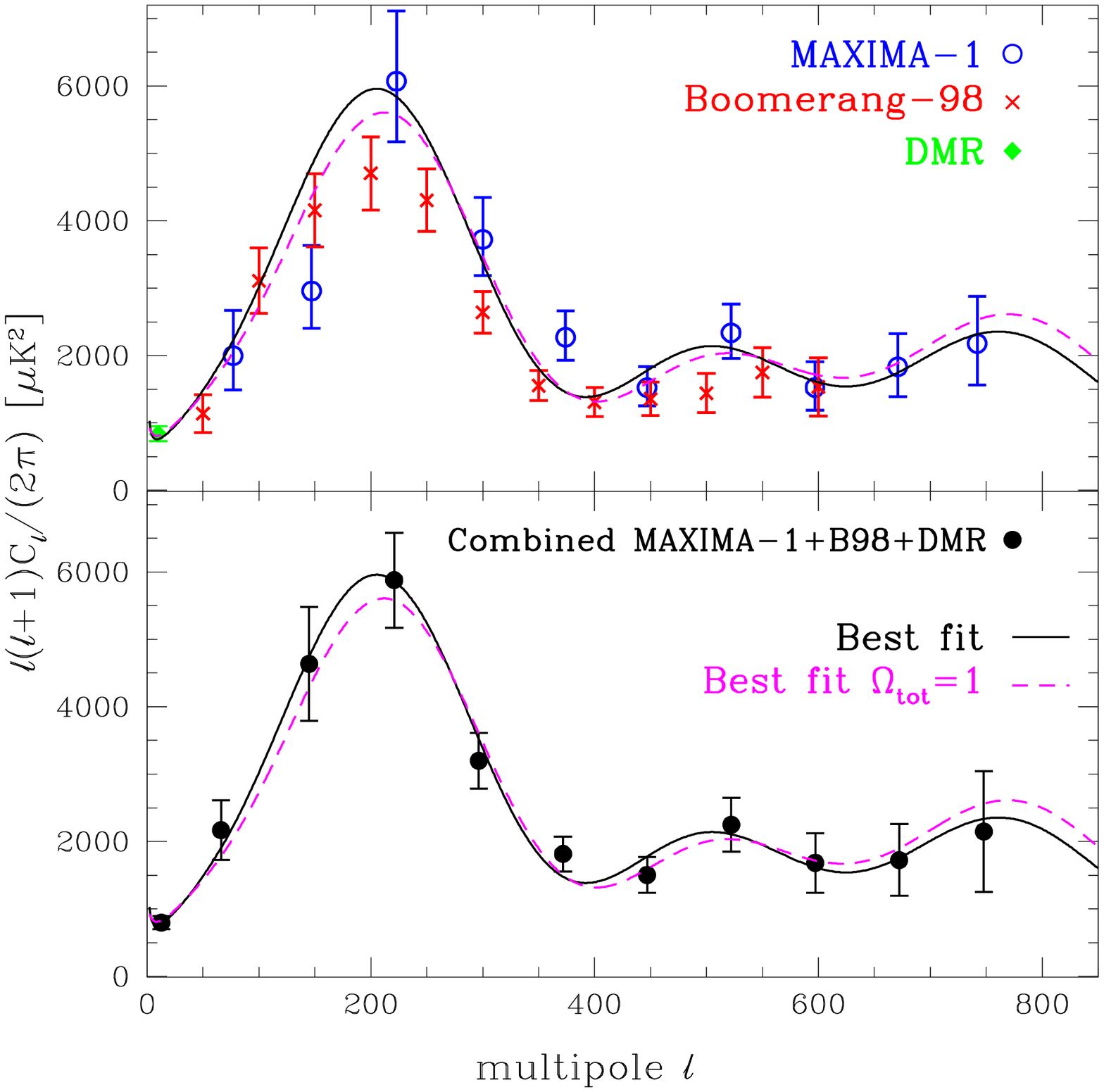,width=5in}}
\fcaption{Data points on the CMB temperature power spectrum
         obtained individually by BOOMERanG and MAXIMA, as well
         as the points obtained from a joint analysis of the two 
         data sets.  The data indicate unequivocally a peak at
         $\ell\sim200$ and are beginning to show the outline of
         a second peak at $\ell\sim500$.  From
         Jaffe et al.\cite{Jafetal00}}
\label{fig:boommax}
\end{figure}

Recent experiments have sought to determine the power spectrum
in the range $50\lsim \ell \lsim 1000$, as
structure-formation theories predict a series of bumps in this
regime that arise as consequences of oscillations in the
baryon-photon fluid in the era before recombination (as
indicated by the curves in Fig. \ref{fig:boommax}).  The
rich structure in these peaks, which can be characterized, e.g., 
by the precise heights and widths of the peaks, their locations in
$\ell$, and the heights of the troughs between the peaks, 
depends in detail on the values of several classical
cosmological parameters, such as the baryon density $\Omega_b$
(in units of the critical density), Hubble constant $h$ (in
units of 100 km/sec/Mpc), matter density $\Omega_m$, and
cosmological constant $\Omega_\Lambda$; on structure-formation
parameters such as the amplitude and spectral index of
primordial perturbations\cite{Junetal96a,Junetal96b}; and on
the character of primordial
perturbations (e.g., adiabatic, isocurvature, or
topological-defects products).  In
particular, the location of the first peak depends primarily on
the geometry of the Universe (parameterized by the {\it total}
density $\Omega_{\rm tot}$), and only secondarily on the other cosmological
parameters\cite{KamSpeSug94}.  If the Universe is flat, the
first peak is expected
to occur at $\ell\sim200$, while if the Universe has a
matter density $\Omega_m\sim0.3$ (as dynamical measurements
indicate) but is open (no cosmological constant), 
then the first peak should be at $\ell\sim500$.

Experiments that measure the power spectrum in the regime
$50\lsim \ell  \lsim 1000$ require high sensitivity to
detect the CMB temperature variations of roughly one part in
100,000, and they require subdegree angular resolution.  Within the past
year, the first high--signal-to-noise high-angular-resolution
maps of the CMB have been published by the BOOMERanG\cite{deBetal00} 
and MAXIMA\cite{Hanetal00} collaborations.  The results of a joint
analysis\cite{Jafetal00} of the data from both BOOMERANG and
MAXIMA  are shown in Fig. \ref{fig:boommax}.
The data show a 
peak at $\ell\sim200$ which provides very strong evidence that
the Universe is flat (earlier measurements by the TOCO
collaboration\cite{Miletal99} also indicated a first peak at
$\ell\sim200$, but with lower signal-to-noise).  The peak
structure is also very
consistent with growth of large-scale structure from a nearly
scale-invariant spectrum of primordial adiabatic perturbations
and very {\it in}consistent with isocurvature or topological-defect
alternatives.  The peak structure indicated in
Fig. \ref{fig:boommax} is also beginning to provide valuable
information about the values of other cosmological
parameters\cite{Lanetal00,Baletal00}.

\section{Inflation and Gravitational Waves}
\noindent
The flatness of the Universe and adiabatic perturbations
suggest that we are on the right track with
inflation\cite{Gut81,Lin82a,AlbSte82}, a period
of accelerated expansion in the very early Universe driven by
the vacuum energy associated with some new ultra-high-energy
physics.  In order to solve the horizon problem for which it was 
initially proposed, inflation predicts that the Universe is
flat. Moreover, shortly after inflation was proposed, it was
realized that vacuum fluctuations in the inflaton (the scalar
field responsible for inflation) would produce a
nearly-scale-invariant spectrum of adiabatic
perturbations\cite{GutPi82,Haw82,Lin82b,Sta82,BarSteTur83}.
With the advent of these CMB tests, inflation has now had several
opportunities to fail empirically, but it has not.
Conservatively, these successes are at least suggestive and
warrant further tests of inflation.

Perhaps the most promising avenue toward further tests of
inflation is the gravitational-wave background.  In addition to
predicting a flat Universe with adiabatic perturbations,
inflation also predicts that quantum fluctuations in the
spacetime metric during inflation would give rise to a
stochastic gravitational-wave background with a
nearly-scale-invariant spectrum\cite{AbbWis84}.  Quantum
fluctuations in the spacetime metric can only be affected by
gravitational effects which are quantified completely during
inflation by the expansion rate $H_{\rm infl}$.  This is related
through the Friedmann equation to
the vacuum-energy density $V$ during inflation, $H_{\rm infl}^2=8\pi V/(3
m_{\rm Pl}^2)$, where $m_{\rm Pl}$ is the Planck mass ($m_{\rm
Pl}^{-2}=G$, Newton's constant, in particle-physics units
$\hbar=c=1$).  Thus, the amplitude of the gravitational-wave
background is fixed entirely by the vacuum-energy density during 
inflation, which itself should be proportional to the fourth
power of the energy scale $E_{\rm infl}$ of the new physics
responsible for inflation.  The spectrum of gravitational waves
depends on the particular inflationary model, but in most models
(and certainly in the simplest inflationary models), it is
likely to be very close to scale invariant.

\begin{figure}[tbp]
\centerline{\psfig{file=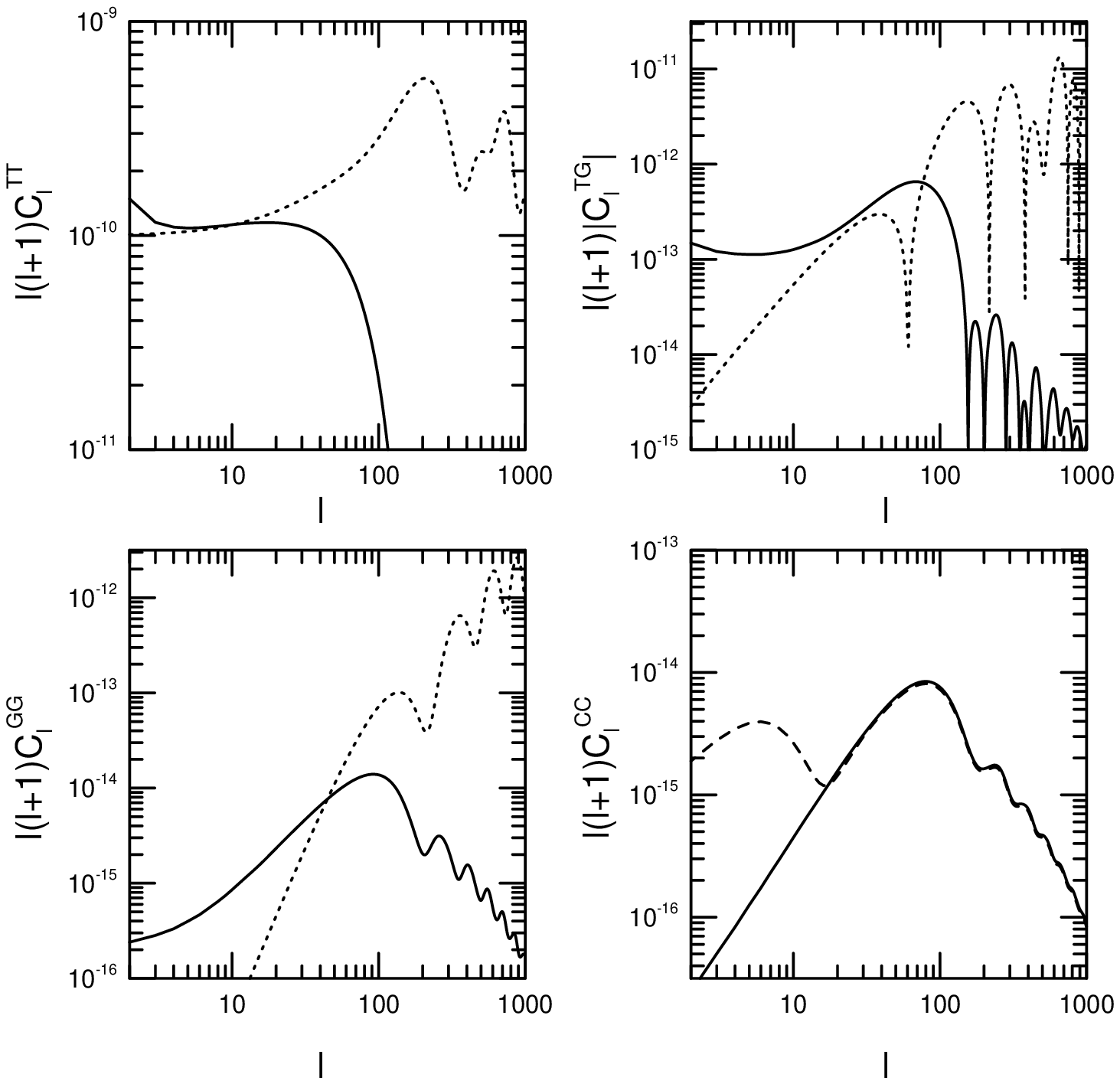,width=6in}}
\fcaption{Temperature and polarization power spectra from density 
         perturbations (dotted curves) and gravitational waves
         (solid curves).  The absence of a dotted curve for the
         CC (lower right-hand) panel is due to the fact that
         density perturbations do not produce a curl component.
         The solid curves show predictions for a model in which
         there is no reionization.  More realistically some
         fraction $\tau\sim0.1$ of the CMB photons will have
         re-scattered from reionized gas, and this will generate 
         additional polarization power at large angles, as
         indicated by the dashed curve in the CC panel.}
\label{fig:cls}
\end{figure}

These gravitational waves will produce temperature fluctuations
at large angles ($\ell \lsim1100$) in the CMB (as shown in
Fig. \ref{fig:cls}).  The amplitude of their contribution to the
CMB temperature quadrupole can be written\cite{KamKos99}
\begin{equation}
  {\cal T} \equiv 6\, C_2^{{\rm TT},{\rm tensor}}= 9.2 {V
     \over m_{\rm Pl}^4},
\end{equation}
(where ``tensor'' refers to gravitational waves, as they are
tensor perturbations to the spacetime metric and ``TT'' refers
to the temperature quadrupole).  Since the quadrupole measured
by {\sl COBE}, $C_2^{\rm TT}=(1.0\pm0.1)\times10^{-10}$, is most 
generally due to some combination of density perturbations and
gravitational waves, we already have an important constraint to
the energy scale of inflation: $V^{1/4}\lsim 2\times 10^{16}$
GeV.\cite{KamKos99,ZibScoWhi99}

\section{Gravitational Waves and Polarization}
\noindent
But how can we go further?  One might think that improved
temperature maps could be used to measure the power spectrum
well enough to distinguish the relative contributions of the
gravitational-wave and density-perturbation power spectra
indicated in the upper left panel of Fig. \ref{fig:cls}.
However, the precision with which the power spectrum can be
measured is limited even in an ideal experiment by cosmic
variance, the sample variance due to the fact that we have only
$2\ell+1$ independent modes with which to measure each
$C_\ell$.

Instead, progress can be made with the polarization of the CMB.
In addition to producing temperature fluctuations, both
gravitational waves and density perturbations will produce
linear polarization in the CMB, and the polarization patterns
produced by each differ.  
This can be quantified with a harmonic decomposition of the
polarization field.  The linear-polarization state of the CMB in 
a direction $\hatn$ can be described by a symmetric
trace-free $2\times2$ tensor,
\begin{equation}
  {\cal P}_{ab}(\hatn)={1 \over 2} \left( \begin{array}{cc}
   \vphantom{1\over 2}Q(\hatn) & -U(\hatn) \sin\theta \\
   -U(\hatn)\sin\theta & -Q(\hatn)\sin^2\theta \\
   \end{array} \right),
\label{eq:whatPis}
\end{equation}
where the subscripts $ab$ are tensor indices, and $Q(\hatn)$ and
$U(\hatn)$ are the Stokes parameters.  Just as the temperature
map can be expanded in terms of spherical harmonics, the
polarization tensor can be
expanded,\cite{KamKosSte97a,KamKosSte97b,SelZal97,ZalSel97}
\begin{equation}
      {{\cal P}_{ab}(\hatn)\over T_0} =
      \sum_{lm} \left[ a_{(lm)}^{{\rm G}}Y_{(lm)ab}^{{\rm
      G}}(\hatn) +a_{(lm)}^{{\rm C}}Y_{(lm)ab}^{{\rm C}}(\hatn)
      \right],
\label{eq:Pexpansion}
\end{equation}
in terms of tensor spherical harmonics, $Y_{(lm)ab}^{\rm G}$
and $Y_{(lm)ab}^{\rm C}$.  It is well known that a vector field
can be decomposed into a curl and a curl-free (gradient)
part.  Similarly, a $2\times2$ symmetric traceless
tensor field can be decomposed into a tensor
analogue of a curl and a gradient part; the $Y_{(lm)ab}^{\rm G}$
and $Y_{(lm)ab}^{\rm C}$ form a complete orthonormal basis for the
``gradient'' (i.e., curl-free) and ``curl'' components of the
tensor field, respectively.\footnote{Our G and C are sometimes
referred to as the ``scalar'' and ``pseudo-scalar''
components\cite{Ste96}, respectively, or with slightly different 
normalization as E and B modes\cite{ZalSel97} (although these
should not be confused with the
radiation's electric- and magnetic-field vectors).}  Lengthy but
digestible expressions for the
$Y_{(lm)ab}^{\rm G}$ and $Y_{(lm)ab}^{\rm C}$ are given in terms 
of derivatives of spherical harmonics and also in terms of
Legendre functions in Kamionkowski et al.\citelow{KamKosSte97b}\ 
The mode amplitudes in Eq. (\ref{eq:Pexpansion}) are given by
\begin{eqnarray}
     a^{\rm G}_{(lm)}&=&{1\over T_0}\int d\hatn\,{\cal P}_{ab}(\hatn)\, 
                                         Y_{(lm)}^{{\rm G}
                                         \,ab\, *}(\hatn),\cr 
     a^{\rm C}_{(lm)}&=&{1\over T_0}\int d\hatn\,{\cal P}_{ab}(\hatn)\,
                                          Y_{(lm)}^{{\rm C} \,
                                          ab\, *}(\hatn), 
\label{eq:Amplitudes}
\end{eqnarray}
which can be derived from the orthonormality properties of these 
tensor harmonics\cite{KamKosSte97b}.
Thus, given a polarization map ${\cal P}_{ab}(\hatn)$, the G and 
C components can be isolated by first carrying out the
transformations in Eq. (\ref{eq:Amplitudes}) to the $a^{\rm
G}_{(lm)}$ and $a^{\rm C}_{(lm)}$, and then summing over the
first term on the right-hand side of Eq. (\ref{eq:Pexpansion})
to get the G component and over the second term to get the C
component. 
(In practice, a full likelihood formalism would be used to determine the
spectra in the presence of anisotropic, correlated noise, astrophysical
foregrounds, and incomplete sky coverage.)

The two-point statistics of the combined
temperature/polarization (T/P) map are specified completely by
the six power spectra $C_\ell^{{\rm X}{\rm X}'}$
for ${\rm X},{\rm X}' = \{{\rm T,G,C}\}$.  Parity invariance
demands that $C_\ell^{\rm TC}=C_\ell^{\rm GC}=0$ (unless the physics
that gives rise to CMB fluctuations is parity
breaking\cite{LueWanKam98,Lep98}).  Therefore, the statistics of
the CMB
temperature-polarization map are completely specified by the
four sets of moments: $C_\ell^{\rm TT}$, $C_\ell^{\rm TG}$, $C_\ell^{\rm
GG}$, and $C_\ell^{\rm CC}$.  

Both density perturbations and gravitational waves will produce
a gradient component in the polarization.  However, only
gravitational waves will produce a curl
component\cite{KamKosSte97a,SelZal97} (but see below).  The curl
component
thus provides a model-independent probe of the gravitational-wave
background, and it is thus the CMB polarization component that
we focus on here.

\section{Detectability of the Curl Component}
\noindent
If our goal is to detect the polarization signature of
gravitational waves, what is the optimum experiment?  What is
the ideal angular resolution and survey size?  What instrumental 
sensitivity is required?  This article will address these
questions (although fall a bit short of providing a complete
answer).

If we are interested only in the gravitational-wave signature,
we can focus on the model-independent curl component of the
polarization produced by gravitational waves.  In this article,
we summarize work reported in Jaffe et al.,\cite{JafKamWan00}\ a
paper that extends the work of
Kamionkowski and Kosowsky\cite{KamKos98} and Lesgourgues et
al.\cite{Lesetal99}.\footnote{There is also
related work in Kinney\cite{Kin98}, Zaldarriaga et
al.\cite{ZalSelSpe97}, and Copeland et al.\cite{Lidetal97b} in
which it is determined how accurately various cosmological and
inflationary parameters can be determined in case of a positive
detection.  Magueijo and Hobson\cite{MagHob97,HobMag96} presented arguments
regarding partial-sky coverage for temperature maps analogous to
those for polarization maps presented here.}
We ask, what is the smallest amplitude of a curl component from
an inflationary gravitational-wave background that could be
distinguished from the null hypothesis of no curl component by
an experiment that maps the polarization over some fraction of the sky
with a given angular resolution and instrumental noise?  
If an experiment concentrates on a smaller
region of sky, then several things happen that affect the
sensitivity: (1) information from modes with
$\ell\lsim180/\theta$ (where $\theta^2$ is the area on the
sky mapped) is lost;\footnote{This is not strictly true. In
principle, as usual in Fourier analysis, less sky coverage
merely limits the independent modes one can measure to have a
spacing of $\delta l\gsim180/\theta$. In practice,
instrumental effects (detector drifts; ``1/f'' noise) will
render the smallest of these bins unobservable.}  (2) the sample
variance is increased; (3) the noise per pixel is decreased
since more time can be spent integrating on this smaller patch
of the sky.

More concretely, suppose we hypothesize that there is a C component of
the polarization with a power spectrum that has the $\ell$ dependence
expected from inflation, as shown in Fig.~\ref{fig:cls},
but an unknown amplitude ${\cal T}$.
We can predict the size of the error that we will obtain from the
ensemble average of the curvature of the likelihood function (also known 
as the Fisher matrix)\cite{Junetal96a,Junetal96b}. In such a
likelihood analysis, the expected error on the
gravitational-wave amplitude ${\cal T}$ will be $\sigma_{\cal
T}$, where
\begin{equation}
     {1\over \sigma_{\cal T}^2} = \sum_\ell \left( { \partial C_\ell^{\rm CC}
     \over \partial {\cal T}} \right)^2 {1\over
     (\sigma_\ell^{\rm CC})^2}.
\label{simplest}
\end{equation}
Here, the $\sigma_\ell^{{\rm CC}}$ are the expected errors at
individual $\ell$ for each $C_\ell^{\rm CC}$ multipole
moments.  These are given by (cf., Kamionkowski et al.\cite{KamKosSte97b}) 
\begin{equation}
  \label{eq:sigl}
  \sigma^{\rm CC}_\ell = \sqrt{2\over f_{\rm sky}(2\ell+1)}
  \left(C_\ell^{\rm CC} + f_{\rm sky} w^{-1} B_\ell^{-2}\right),
\end{equation}
where $w^{-1}=4\pi s^2 /(t_{\rm pix} N_{\rm pix}T_0^2)$ is the
variance (inverse weight) per unit area on
the sky, $f_{\rm sky}$ is the fraction of the sky observed, and
$t_{\rm pix}$ is the time spent observing each of the $N_{\rm pix}$
pixels. The detector sensitivity is $s$ and the average sky temperature
is $T_0=2.73\,\mu{\rm K}$ (and hence the $C_\ell^{\rm CC}$ are measured
in units that have been scaled by $T_0$). The inverse weight for a full-sky
observation is $w^{-1}=2.14\times10^{-15} t_{\rm yr}^{-1}(s/200\,
\mu{\rm K}\, \sqrt{\rm sec})^2$ with $t_{\rm yr}$ the total observing
time in years. Finally, $B_\ell$ is the experimental beam, which for a
Gaussian is $B_\ell=e^{-\ell^2 \sigma_\theta^2/2}$.  We assume all
detectors are polarized.  

The error to $C_\ell^{\rm CC}$ has two terms, one proportional
to $C_\ell^{\rm CC}$ (the {\em sample variance}), and another
proportional to $w^{-1}$ (the {\em noise variance}). 
There are several complications to note when considering these formulae:
1) We never have access to the actual $C_\ell^{\rm CC}$, but only to
some estimate of the spectra; 2) the expressions only deal approximately
with the effect of partial sky coverage; and 3) the actual likelihood
function can be considerably non-Gaussian, so the expressions above do
not really refer to ``$1\sigma$ confidence limits.''

Here, we are interested in the detectability of the curl component;
that is, what is the smallest gravitational-wave amplitude that we
could confidently differentiate from zero?  Toy problems and
experience give us an approximate rule of thumb: the signal is
detectable when it can be differentiated from the ``null
hypothesis'' of $C_\ell^{\rm CC}=0$.  
Thus, the $\ell$ component of the gravitational-wave signal is
detectable if its amplitude is greater than
\begin{equation}
     \sigma_\ell^{\rm CC} = \sqrt{2/(2\ell+1)} f_{\rm sky}^{1/2} 
     w^{-1} e^{\ell^2 \sigma_b^2}.
\label{CCnoise}
\end{equation}
We then estimate the smallest gravitational-wave amplitude
${\cal T}$ that can be distinguished from zero (at ``1 sigma'')
by using Eq. (\ref{simplest}) with the null hypothesis $C_\ell^{\rm
CC}=0$.  Putting it all together, the smallest detectable
gravitational-wave amplitude (scaled by the largest consistent
with {\sl COBE}) is
\begin{equation}
     {\sigma_{\cal T} \over {\cal T}} \simeq
     1.47\times10^{-17}\, t_{\rm yr} \, \left( {s \over 200\,
     \mu {\rm K}\sqrt{\rm sec}} \right)^2 \, \left({\theta \over 
     {\rm deg}} \right) \, \Sigma_\theta^{-1/2},
\end{equation}
where
\begin{equation}
     \Sigma_\theta = \sum_{\ell\geq (180/\theta)} (2\ell+1) \left(
     C_\ell^{\rm CC} \right)^2 e^{-2\ell^2\sigma_b^2}.
\label{summand}
\end{equation}

\begin{figure}[tbp]
\centerline{\psfig{file=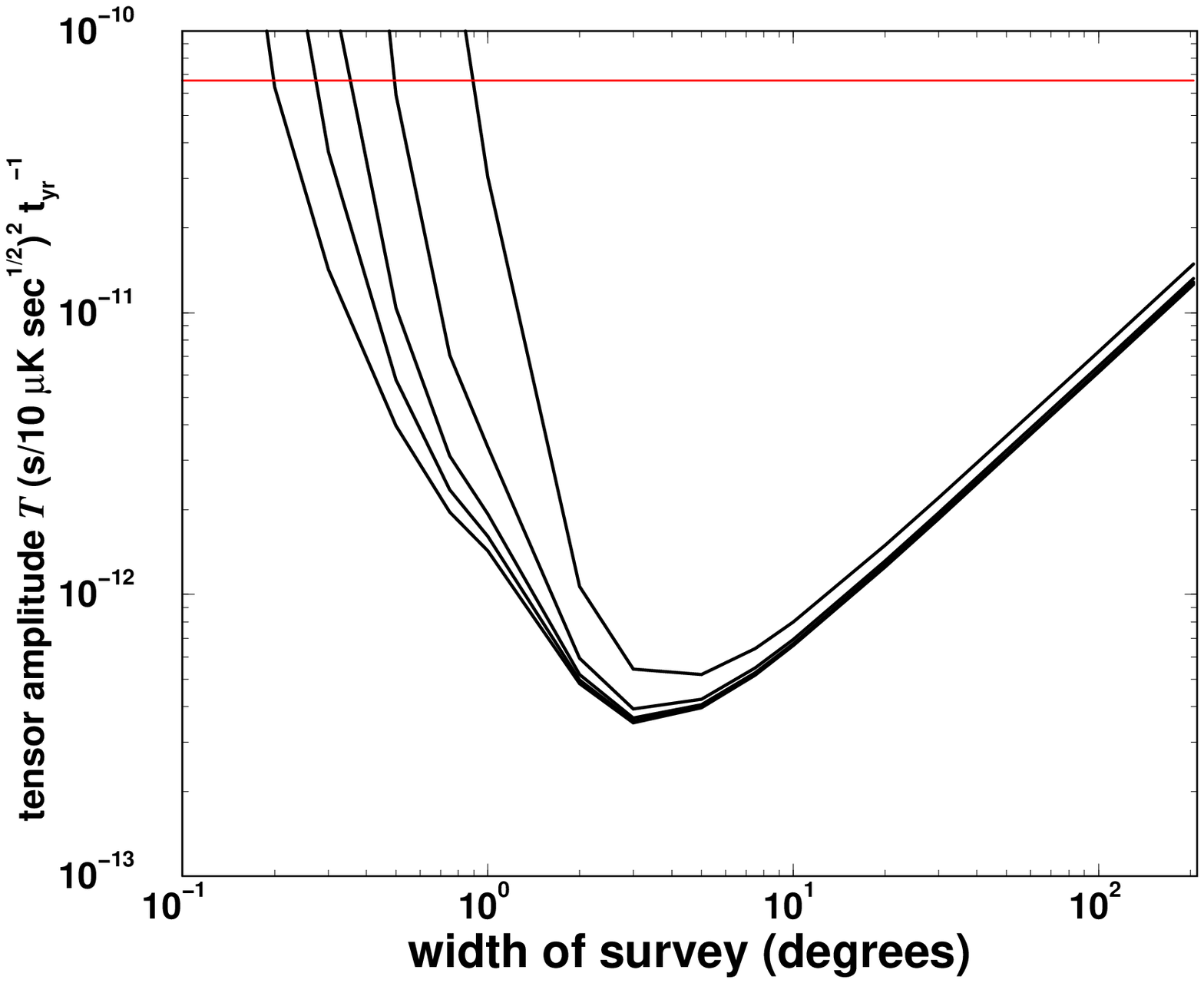,width=5in}}
\fcaption{The smallest gravitational-wave (tensor) amplitude ${\cal T}$ that
         could be detected at $3\sigma$ with an experiment with a detector
         sensitivity $s = 10\,\mu {\rm K} \sqrt{\rm sec}$ that
         runs for one year and maps a square region of the sky
         of a given width.  The result scales with the square of 
         the detector sensitivity and inversely with the
         duration of the experiment.  The curves are (from top
         to bottom) for fwhm beamwidths of 1, 0.5, 0.3, 0.2, and
         0.1 degrees.  The horizontal line shows the upper limit 
         to the gravitational-wave amplitude from {\sl COBE}.}
\label{fig:tensorsens}
\end{figure}

Results of the calculation are shown in
Fig.~\ref{fig:tensorsens}.  Plotted there is the
smallest gravitational-wave amplitude ${\cal T}$
detectable at $3\sigma$ by an experiment with a detector sensitivity
$s=10\,\mu {\rm K} \sqrt{\rm sec}$ that maps a square region of the sky
over a year with a given beamwidth.  The horizontal line shows the upper
limit to the gravitational-wave amplitude from {\sl COBE}.  The
curves are (from top to
bottom) for fwhm beamwidths of 1, 0.5, 0.3, 0.2, 0.1, and 0.05 degrees.
The results scale with the square of the detector sensitivity and
inversely to the duration of the experiment.

The sensitivity to the gravitational-wave signal is a little better
with an 0.5-degree beam than with a 1-degree beam, but even
better angular resolution does not improve the sensitivity
much.  And with a resolution of 0.5 degrees or better, the best
survey size for detecting this gravitational-wave signal is about 3 to 5
degrees.  If such a fraction of the sky is surveyed, the
sensitivity to a gravitational-wave signal (rms) will be about 30 times
better than with a full-sky survey with the same detector
sensitivity and duration (and thus 30 times better than
indicated in Kamionkowski and Kosowsky\cite{KamKos98,KamKos99}.
Thus, a balloon
experiment with the same detector sensitivity as MAP could in
principle detect the same gravitational-wave amplitude in a few weeks that
MAP would in a year.  (A width of 200 degrees corresponds to
full-sky coverage.)

\begin{figure}[tbp]
\centerline{\psfig{file=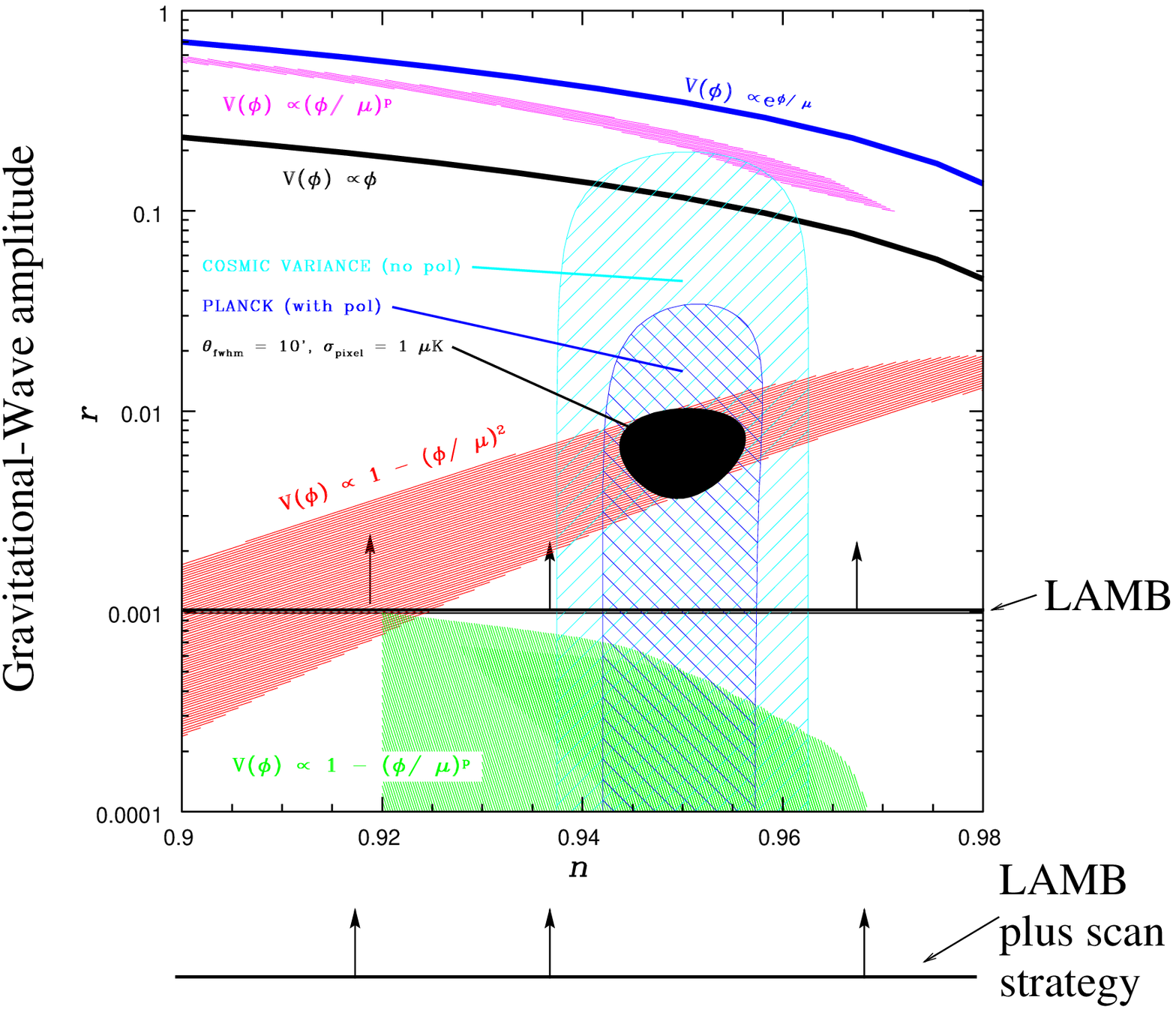,width=5in}}
\fcaption{Regions in the $r$-$n$ parameter space space occupied
         by various inflationary models, as well as those
         regions that could be detected by various CMB
         experiments.  Here $r$ measures the gravitational-wave
         amplitude (or alternatively, the energy scale of
         inflation) and $n$ the spectral index for primordial
         perturbations.  Adapted from Kinney\cite{Kin98}.  See
         text for more details.}
\label{fig:kinney}
\end{figure}

Since the gravitational-wave amplitude is related to the energy
scale of inflation, Fig.~\ref{fig:tensorsens} 
determines the inflationary energy scale accessible with any
given experiment.  Some indication of the range of inflationary
models that can be probed with past, current, and future
experiments is provided in Fig. \ref{fig:kinney}, adapted from
Kinney\cite{Kin98}.  The parameter $r$ ($y$ axis) increases with
increasing gravitational-wave amplitude, or alternatively, with
the energy scale of inflation.  The shaded
regions show the where the predictions for various classes of
inflationary models (e.g., exponential, power-law, etc; for more 
details see Kinney et al.\cite{KinDodKol97,Kin98}).  The scored region
labeled ``COSMIC VARIANCE (no pol)'' shows the region of
parameter space that would be consistent with a null
search for gravitational waves with{\it out} polarization,
while that labeled ``PLANCK (with pol)'' shows regions of
parameter space that would be consistent with a null search for
the curl component in the Planck satellite\cite{Planck}, an ESA 
CMB mission scheduled for launch in 2007 (in both cases, it is
assumed that $n=0.95$).

How much could the sensitivity be improved with a post-Planck
dedicated polarization experiment?  Achieving the Planck
sensitivity will be no small feat for experiment, and
improvements will require considerable ingenuity.  Still, there
are prospects for improvements.  Very conservatively, a
factor-of-3 improvement over Planck's detector sensitivity is
plausible.  The dark ellipse labeled ``$\theta_{\rm
fwhm}=10^\circ$, $\sigma_{\rm pixel}=1 \mu$K'' is the error
ellipse that could be obtained by a putative experiment with
roughly a factor-of-3 improvement to Planck's detector
sensitivity, assuming that the true gravitational-wave
amplitude and spectral index lie at the center of that ellipse.

There are good reasons to believe that technological
developments in the next few years may allow further
improvements in detector sensitivity, perhaps of an order of 
magnitude over that achieved in Planck.  As an example,
we mention LAMB (Large-Format Array of Microwave
Bolometers)\cite{LAMB}, a new detector concept that would allow
roughly an order-of-magnitude improvement over Planck's
detector sensitivity with a much smaller instrument.  If we
assume that such a detector can be developed and flown in an
all-sky survey, the factor-of-ten improvement would allow us to
access the regions of parameter space that lie above the line
labeled ``LAMB'' in Fig. \ref{fig:kinney}.  If this experiment
additionally spent its time surveying a smaller region of the
sky, then the regions of inflationary parameter space that could 
be accessed would be those that lie above the line labeled
``LAMB plus scan strategy.''

\section{Conclusions}
\noindent
We have carried out calculations that will help assess the prospects for
detection of the curl component of the polarization with various
experiments.  Our results can be used to forecast the
signal-to-noise for the gravitational-wave signal in an
experiment of given sky coverage, angular resolution, and
instrumental noise.  Of course, the ``theoretical''
factors considered here must be weighed in tandem with those that
involve foreground subtraction and experimental logistics in the design
or evaluation of any particular experiment. These usually
encourage increasing the signal-to-noise and sky coverage to
better isolate experimental systematics.

In contrast to temperature anisotropies which show power on all scales
[i.e., $\ell(\ell+1)C_\ell\sim{\rm const}$], the polarization
power peaks strongly at higher $\ell$. Hence the signal-to-noise in a
polarization experiment of fixed flight time and instrumental
sensitivity may be improved by surveying a smaller region of sky, unlike
the case for temperature-anisotropy experiments.  The ideal survey for
detecting the curl component from gravitational waves is of order 2--5
degrees, and the sensitivity is not improved much for angular
resolutions smaller than 0.2 degrees.  An experiment with this
ideal sky coverage could improve on the sensitivity to
gravitational waves of a full-sky experiment by roughly a factor 
of 30.  When coupled with realistic forecasts for improvements
in detector sensitivity, we find that an experiment that accesses a
very good fraction of the inflationary parameter space
(specifically, most of the inflationary parameter space
associated with grand unification) is conceivable in the
not-too-distant future.

Before closing, we should note that secondary (in the
density-perturbation amplitude) effects such as weak
gravitational lensing\cite{ZalSel98} or re-scattering of CMB
photons from reionized gas\cite{Hu99} may lead to the 
production of a curl component in the CMB, even in the absence
of gravitational waves.  However, these secondary effects
should be distinguishable from those of gravitational waves, as
they produce a curl component primarily at angular scales much 
smaller than those at which the gravitational-wave signal should 
show up.  Of course, an angular resolution better than that we
have suggested here and a survey area a bit larger than we have
suggested here may be required to distinguish the 
gravitational-wave signal from these other sources of a curl
component (as well as from foregrounds).
A more complete assessment of the impact of these
secondary effects on the detectability of gravitational waves is 
now underway\cite{KesKam00}.

Finally, the CMB polarization will be useful for a wide variety
of other purposes in cosmology.  For example, detection, and
ultimately mapping, of the polarization will help isolate the
peculiar velocity at the surface of last scatter\cite{ZalHar95},
constrain the ionization history of the
Universe\cite{Zal97}, determine the nature of primordial
perturbations\cite{Kos98,SpeZal97}, probe primordial magnetic
fields\cite{KosLoe96,HarHayZal96,ScaFer97} and cosmological
parity violation\cite{LueWanKam98,Lep98}, and maybe more (see,
e.g., Kamionkowski and Kosowsky\cite{KamKos99} for a recent review).  

\nonumsection{Acknowledgments}
\noindent
MK was supported in part by NSF AST-0096023, NASA NAG5-8506, and
DoE DE-FG03-92-ER40701.  AHJ was supported by NSF KDI grant
9872979 and NASA LTSA grant NAG5-6552.

\end{document}